\documentclass{article}
\usepackage[utf8]{inputenc} 
\usepackage{amssymb, amsmath, amsthm, graphicx, subfigure}
\usepackage[dvipsnames]{xcolor}
\usepackage[colorlinks=true,linktoc=page]{hyperref}
\usepackage[margin=1in]{geometry}
\usepackage{mathtools}
\usepackage{tikz}
\usepackage{float}
\usepackage{cleveref}
\usepackage{enumitem}
\usepackage{xspace}

\mathtoolsset{showonlyrefs=true}

\theoremstyle{plain}
\newtheorem*{structuralrestatement}{Theorem~\ref{thm:structural}}

\newcommand{\intervalstructurestatement}{%
    For all interval graphs $G = (V, E)$ and $\e > 0$, there is a set
    of edges $F \subseteq E$ such that $|F| \leq \e |E|$ and
    $G' = (V, E \setminus F)$ has rank-width at most
    $\mathcal O(\e^{-5} \log \e^{-1})$.
    Further, $F$ can be computed in polynomial time.%
}

\newtheorem*{ptasrestatement}{Theorem~\ref{thm:ptas}}

\newcommand{\rankwidthptasstatement}{%
    Given an instance of \csp with constraint graph $G = (V, E)$, a
    rank-decomposition of $G$ of width at most $r$, and $\e > 0$,
    there is a randomized $n^{\mathcal O(r\e^{-2})}$-time algorithm
    that returns a labeling $\sigma : V \to \zo$ with
    $\mathbb E[\val{\sigma}] \geq (1-\e)\opt$.%
}
\newtheorem{theorem}{Theorem}[section]
\newtheorem{lemma}[theorem]{Lemma}
\newtheorem{corollary}[theorem]{Corollary}

\newtheorem{observation}[theorem]{Observation}

\theoremstyle{definition}
\newtheorem{definition}[theorem]{Definition}

\theoremstyle{remark}

\def\set#1{\{ #1 \}}
\def\EMPH#1{\emph{\textcolor{BrickRed} {#1}}}

\newcommand{\val}[1]{\operatorname{Value(#1)}}
\newcommand{\opt}{\operatorname{OPT}}
\newcommand{\csp}{\textsc{Boolean Max-2-CSP}\xspace}
\newcommand{\zo}{\set{0, 1}}

\newcommand{\e}{\varepsilon}

\hypersetup{colorlinks,
    linkcolor=blue,
    citecolor=ForestGreen,  
    urlcolor=Mahogany,
}

\title{A PTAS for Constraint Satisfaction Problems on Interval Graphs and Bounded Rank-Width Graphs }
\author{Ian DeHaan\thanks{University of Michigan. \texttt{idehaan@umich.edu} }\and Euiwoong Lee\thanks{University of Michigan. \texttt{euiwoong@umich.edu} }}
\date{}

\begin{document}

\maketitle

\begin{abstract}
    We present a polynomial-time approximation scheme (PTAS) for all Boolean arity-two maximum constraint satisfaction problems (\csp) where the constraint graph is either interval or has bounded rank-width.
    This class of problems includes, for example, \textsc{Max-Cut}, \textsc{Max-DiCut}, \textsc{Max-2-Lin(2)}, and \textsc{Max-2-SAT}.
    In particular, this settles the approximability of \textsc{Max-Cut} on interval graphs.
    To achieve these results, we show the following.
    \begin{enumerate}
        \item For any $\e > 0$, there is an $n^{\mathcal O(r \e^{-2})}$-time $(1 - \e$)-approximation for \csp on graphs of rank-width at most $r$.
        \item There is a polynomial-time algorithm that, for any interval graph $G = (V, E)$ and $\e > 0$, finds a subgraph $H = (V, F)$ of $G$ such that $|F| \geq (1 - \e) |E|$ and $H$ has rank-width at most $\operatorname{poly}(\e^{-1})$.
    \end{enumerate}
    Our techniques additionally extend to all bounded-alphabet arity-two max CSPs.
    
\end{abstract}

\section{Introduction}

An instance of \csp consists of a simple constraint graph $G = (V, E)$ and a non-zero constraint \mbox{$f_{uv} : \zo^2 \rightarrow \zo$} for each edge $uv \in E$.
The goal is to find a labeling $\sigma : V \to \zo$ maximizing the number of constraints satisfied:
\[
    \val{\sigma} := \sum_{uv \in E} f_{uv}(\sigma(u), \sigma(v)).
\]
This problem is general enough to capture many classic optimization problems.
For example, one may express \textsc{Max-Cut} by setting $f_{uv}(a, b) = 1$ iff $a \neq b$ for all edges $uv$, or \textsc{Max-DiCut} by setting $f_{uv}(a, b) = 1$ iff $a = 0$ and $b = 1$ for all directed edges $u \rightarrow v$.

We say that an algorithm is a \EMPH{$c$-approximation} for \csp if it produces some labeling $\sigma : V \to \zo$ such that $\val{\sigma} \geq c \opt$, where $\opt$ is the value of the optimal labeling.
A problem admits a \EMPH{polynomial-time approximation scheme} (PTAS) if there exists a polynomial-time $(1 - \e)$-approximation for all $\e > 0$.
Our paper presents two structural restrictions on the constraint graph $G$ which allow for a PTAS --- namely, we show that there is a PTAS for \csp on interval graphs and bounded rank-width graphs.

\paragraph{Interval graphs.}
A graph $G = (V, E)$ is \EMPH{interval} if there exists a collection of closed intervals $\set{\mathcal I_v}_{v \in V}$ on the real line such that $uv \in E$ iff $\mathcal I_u$ and $\mathcal I_v$ intersect. %
Interval graphs commonly arise in real-world graph data. 
For example, they often appear as register-conflict graphs, where intervals correspond to ``live ranges'' of variables, or in biology, where they can be used to model subjects such as DNA and food webs \cite{JUNGCK20151,hendren1992register}.

\begin{figure}[h]
    \centering
    \includegraphics[width=0.75\linewidth]{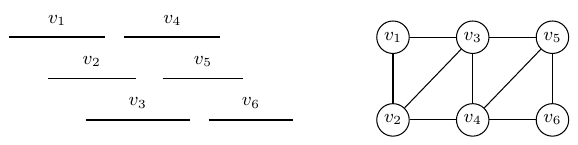}
    \caption{An interval graph. The left figure displays an interval representation of the graph on the right.}
\end{figure}

\paragraph{The rank-width of a graph.}
Rank-width, introduced by Oum and Seymour in 2006 \cite{oum2006approximating}, is a measure of the \emph{complexity} of a graph.
Roughly speaking, a graph has rank-width at most $r$ if it can be recursively cut into balanced parts such that each cut induces a matrix of rank at most $r$.
One can view bounded rank-width graphs as a generalization of bounded tree-width graphs to also include ``simple'' dense graphs.
For example, the tree-width of the complete graph $K_n$ is $n-1$, but has rank-width only $1$.
Bounded rank-width is equivalent to several other graph parameters such as bounded clique-width and bounded module-width.
Rank-width has a distinct advantage over these other parameters in that the rank-width $r$ of a graph can be calculated \emph{exactly} in $\mathcal O(n^3)$-time, albeit with a large constant factor depending on $r$ \cite{hlinveny2008finding}.
See \Cref{subsection:prelims} for a formal definition of rank-width.

\paragraph{Our results.}
Our main results are as follows.
\begin{theorem} \label{thm:intervalptas}
    \csp admits a randomized PTAS on interval graphs.
\end{theorem}
\begin{theorem} \label{thm:rankwidthptas}
    \csp admits a randomized PTAS on bounded rank-width graphs.
\end{theorem}

We show \Cref{thm:rankwidthptas} by presenting a combinatorial divide-and-conquer type algorithm which, in expectation, returns a $(1 - \e)$-approximate solution in time $n^{\mathcal O(r \e^{-2})}$ on graphs of rank-width $r$.
We then show \Cref{thm:intervalptas} through a reduction to \Cref{thm:rankwidthptas}.
In particular, we show that interval graphs are ``nearly'' bounded rank-width --- we can delete an $\e$-fraction of the edges of an interval graph to obtain a bounded rank-width graph.
\begin{theorem} \label{thm:structural}
    \intervalstructurestatement
\end{theorem}

\Cref{thm:intervalptas} then follows immediately from \Cref{thm:rankwidthptas} and \Cref{thm:structural}.
Given an interval graph $G = (V, E)$, we first compute a set $F \subseteq E$ as guaranteed by \Cref{thm:structural} and then run our PTAS from \Cref{thm:rankwidthptas} on the graph $G' = (V, E \setminus F)$. 
The approximation factor follows from noting that $\opt \geq \frac{|E|}{4}$, as a random labeling satisfies each constraint with probability at least $\frac{1}{4}$.

\subsection{Related Prior Work}

\paragraph{CSPs on interval graphs.}
Despite their ubiquity and simple definition, there has been relatively little progress on understanding the approximability of CSPs on interval graphs.
Take \textsc{Max-Cut} as an example, arguably the simplest and most famous constraint satisfaction problem.
Until 2020, \emph{it was not even known that \textsc{Max-Cut} is NP-hard on interval graphs} \cite{adhikary2023complexity}. 
Following the renewed attention to this problem that the hardness result of \cite{adhikary2023complexity} generated, it was shown in 2025 that there is an $(\alpha_{GW} + 10^{-34})$-approximation for \textsc{Max-Cut} on interval graphs \cite{adkl25}.
Here, $\alpha_{GW} \approx 0.878$ is the Goemans-Williamson constant, the optimal approximability of \textsc{Max-Cut} on general graphs \cite{goemans1995improved,khot2007optimal}.
Prior to our work, the approximability of \textsc{Max-Cut} on interval graphs remained a mystery --- should it be close to $1$, close to $\alpha_{GW}$, or somewhere in between?
To our knowledge, there have been no results concerning the approximability of other CSPs on interval graphs prior to our work.

\paragraph{CSPs on bounded rank-width graphs.}
It was previously known that the maximum cut of a bounded rank-width graph can be computed exactly \cite{wanke1994k}.
Additionally, when parameterizing by $r$, there is a fixed-parameter-tractable $(1 - \e)$-approximation for \textsc{Max-Cut} \cite{lampis2014parameterized}.
Notably, both these algorithms work only for \textsc{Max-Cut} and do not readily generalize to varying constraints as in \csp.
Indeed, \csp is NP-hard even on the complete graph, which has rank-width $1$:
Given a \textsc{Max-Cut} instance, assign inequality constraints to its edges and always-satisfied dummy constraints to its non-edges.

\paragraph{CSPs on other structured graphs.}
There are an abundance of avenues through which CSPs on structured graphs can be studied.
The preeminent work in this domain is that of \cite{grohe2001evaluation}, which shows that max CSPs with unbounded alphabet are polynomial-time solvable on a class of graphs if and only if that class has bounded tree-width.
When one turns to approximability, there is substantially less known.
The parameter \emph{treewidth-pliability} is introduced in \cite{romero2021treewidth}, where it is shown that bounded treewidth-pliability is sufficient to obtain a PTAS for approximating the optimal value of a broad class of max CSPs. 
The authors of \cite{romero2021treewidth} additionally conjecture that bounded treewidth-pliability should be necessary to obtain a PTAS for these CSPs.
We note that their class of CSPs is substantially broader than what we consider in this paper --- they relax both our bounded alphabet and arity-two requirements.

Another well-studied class of max CSPs are the \emph{dense CSPs} with bounded alphabet --- for arity-two, these correspond to graphs with $\Omega(n^2)$ edges.
For these instances, one can obtain a PTAS through regularity lemmas \cite{frieze1996regularity}, convex hierarchies \cite{de2007linear,yoshida2014approximation}, and sampling \cite{arora1995polynomial}. 
This sampling approach is a major inspiration for our PTAS on bounded rank-width graphs, which borrows some of the key ideas of \cite{arora1995polynomial} in its design.

\paragraph{Comparison with \cite{ahmadzadehAlimi2026eptas}.}
During preparation of this paper, the authors were made aware of concurrent work regarding \textsc{Max-Cut} on interval graphs.
In a submission to STOC'27, Ahmadzadeh and Alimi show there is an EPTAS for \textsc{Max-Cut} on interval graphs.
They present a structural theorem similar to ours --- they show that interval graphs can be transformed into bounded clique-width graphs after at most $\e |E|$ edge deletions.
Recall that a graph has bounded rank-width if and only if it has bounded clique-width \cite{oum2006approximating}, so their structural theorem is essentially equivalent to \Cref{thm:structural} up to dependence on $\e$.
A randomized EPTAS for \textsc{Max-Cut} on interval graphs then follows from the known PTAS for \textsc{Max-Cut} parameterized by clique-width \cite{lampis2014parameterized}.
However, \cite{ahmadzadehAlimi2026eptas} also presents a completely self-contained deterministic EPTAS.

\subsection{Preliminaries} \label{subsection:prelims}

\paragraph{Graph notation.}
For a graph $G = (V, E)$, we write $n := |V|$ and $m := |E|$ when the graph $G$ we refer to is clear from context.
For $v \in V$, we write $N(v)$ to indicate the neighbors of $v$.
For a subset of vertices $S \subseteq V$, we let $\delta(S) := \{uv \in E \mid u \in S, v \not \in S\}$ be the set of edges crossing the cut defined by $S$. 
We will also often write $G[S]$ to be the subgraph of $G$ induced by $S$ and $E[S] := \{uv \in E \mid u \in S, v \in S\}$ to be the edges contained entirely in $S$.
We write $N(S) := \cup_{v \in S} N(v)$ to indicate the neighbors of $S$.

\paragraph{Rank-width.}
We will now formally define the rank-width of a graph.

\begin{definition}
    For a graph $G = (V, E)$ and $S \subseteq V$, we define \EMPH{$\rho_G(S)$} as the rank over $\mathbb F_2$ of the $|S| \times |V \setminus S|$ matrix $A_S$ which contains a $1$ at entry $(u, v) \in S \times (V \setminus S)$ if $uv \in E$, and $0$ otherwise.
\end{definition}

We refer to $\rho_G(S)$ as the rank of the cut $\delta(S)$.
When the graph $G$ is obvious from context, we will often omit $G$ and write $\rho(S) := \rho_G(S)$.
Note that the rank of a cut is at most the number of edges crossing the cut, but may be quite a bit lower.
For example, every cut of a complete graph has rank at most $1$.
We now define precisely what we mean by recursively cutting a graph into balanced parts with cuts of rank at most $r$.
We say that a tree is \EMPH{sub-cubic} if every vertex has degree at most $3$.

\begin{definition}
    A \EMPH{rank-decomposition} of a graph $G = (V, E)$ consists of a sub-cubic tree $T$ and a bijection between $V$ and the leaves of $T$.
    Each edge of the tree $e \in E(T)$ implies a cut $S_e \subseteq V$ of $G$, where $S_e$ is the set of vertices associated with leaves on one side of $e$. 
    The \EMPH{width} of a rank-decomposition is the maximum of $\rho(S_e)$ over all edges in the tree $e \in E(T)$.
    The \EMPH{rank-width} of $G$ is the minimum width over all rank-decompositions of $G$.
\end{definition}

A simple consequence of this definition is that every graph with at least two vertices and rank-width at most $r$ admits a cut $V_1 \subseteq V$ such that
\[
    \frac{n}{3} \leq |V_1| \leq \frac{2n}{3} \qquad \text{and} \qquad \rho_G(V_1) \leq r.
\]
This cut $V_1$ can be obtained from a width-$r$ rank-decomposition of $G$ by taking the cut associated with some balanced edge of $T$.
This is the core property that our algorithm for \csp on bounded rank-width graphs will rely on.

\section{Overview of Our Techniques}

\subsection{PTAS For Bounded Rank-Width Graphs}

In this subsection, we give a brief overview of the techniques we use to obtain a PTAS for \csp on bounded rank-width graphs.
Recall that an optimal rank-decomposition of a graph with bounded rank-width can be computed in polynomial time \cite{hlinveny2008finding}, so we assume that we are supplied with a rank-decomposition of width $r$.

\begin{theorem} \label{thm:ptas}
    \rankwidthptasstatement
\end{theorem}

\paragraph{The algorithmic framework.}
On input $G = (V, E)$ with rank-width at most $r$, recall that we can use the rank-decomposition to obtain a partition $V_1, V_2 \subseteq V$ with the following properties:
\[
    \frac{n}{3} \leq |V_1| \leq \frac{2n}{3} \qquad \text{and} \qquad \rho(V_1) \leq r.
\]
Intuitively speaking, we wish to recurse on $G[V_1]$ to find a labeling $\sigma_1 : V_1 \to \zo$ and on $G[V_2]$ to find a labeling $\sigma_2 : V_2 \to \zo$ which we can then combine to produce a good labeling for $V$.
The issue with this is that, even if $\sigma_1$ is optimal in $G[V_1]$ and $\sigma_2$ is optimal in $G[V_2]$, they may satisfy very few or none of the constraints of the cut $\delta(V_1)$.
Thus, when we recurse on $G[V_1]$ and $G[V_2]$, we wish to somehow encode \emph{awareness} of $\delta(V_1)$ in the subinstances, so we can obtain labelings $\sigma_1$ and $\sigma_2$ that will satisfy much of $\delta(V_1)$.

This is somewhat difficult to do, as the subinstances then lose their independence from each other.
We will return to this problem later. 
Suppose for now that we are given an optimal labeling restricted to $V_2$.
That is, let $\sigma_{\opt} : V \to \zo$ be an optimal labeling, and suppose we are given the information $\sigma_2 : V_2 \to \zo$ where $\sigma_2(v) = \sigma_{\opt}(v)$ for all $v \in V_2$.
Then we can \emph{encode} the value of constraints in the cut $\delta(V_1)$ by introducing \emph{unary rewards} for the subinstance $G[V_1]$.
Concretely, we introduce unary reward functions $\{h_u\}_{u \in V_1}$ where $h_u : \zo \to \mathbb N$ and set
\[
    h_u(b) := \sum_{uv \in \delta(V_1)} f_{uv}(b, \sigma_2(v)).
\]
Then the objective function for the subinstance defined by $G[V_1]$ becomes
\[
    \text{maximize} \sum_{uv \in E[V_1]} f_{uv}(\sigma_1(u), \sigma_1(v)) + \sum_{u \in V_1} h_u(\sigma_1(u)).
\]
Now if we obtain a (near-)optimal solution $\sigma_1$ to this subinstance, this provides a (near-)optimal solution to the original input when combined with $\sigma_2$.
The core difficulty is now this: how can we find a ``good'' labeling $\sigma_2$ to use for computing $\sigma_1$?

\paragraph{Approximating $\sigma_{\opt}$ via sampling.}
Let $N := N(V_1) \cap V_2$ be the neighbors of $V_1$ in $V_2$.
Note that, to compute $\sigma_1$, we only need to know the labels that $\sigma_{\opt}$ assigns to $N$.
Thus, if $N$ were sufficiently small, we could just \emph{guess} a labeling $\sigma_N : N \to \zo$ by iterating over all $2^{|N|}$ possibilities for $\sigma_N$.
Unfortunately, $N$ may be too large for this to run fast enough.
We will instead attempt to ``approximate'' $N$ by randomly selecting some smaller subset $S \subseteq N$, and inferring information about $\sigma_{\opt}$ from this set.

In particular, we will select, for each vertex $u \in V_1$, a uniformly random subset $S_u \subseteq N(u) \cap V_2$, where $|S_u| = \min\{\e^{-2}, |N(u) \cap V_2|\}$.
We then set $S = \cup_{u \in V_1} S_u$ and guess a labeling $\sigma_S : S \to \zo$ by iterating over all $2^{|S|}$ possibilities, as before.
We will later use the rank of the cut to bound the size of $S$, and thus the runtime of our algorithm --- for now we focus only on correctness.
Now, we calculate the \emph{estimated} unary reward $h_u : \{0, 1\} \to \mathbb R_{\geq 0}$ for $u \in V_1$ as
\[
    h_u(b) := \frac{|N(u) \cap V_2|}{|S_u|}\sum_{v \in S_u} f_{uv} (b, \sigma_S(v)).
\]
We claim that the functions $\{h_u\}_{u \in V_1}$ are ``not too far'' from what they would be if we sampled all neighbors $S = N$.
More formally, fix some optimal labeling $\sigma_{\opt}$ and define
\[
    h_u'(b) := \sum_{uv \in \delta(V_1)} f_{uv}(b, \sigma_{\opt}(v)).
\]
Suppose we are in the branch in which $\sigma_S$ is selected to agree with $\sigma_{\opt}$ on $S$.
Now define 
\[
    \Delta_u := \max_{b \in \{0, 1\}} |h_u(b) - h_u'(b)|
\]
as the maximum difference between $h_u$ and $h_u'$, and let $\Delta := \sum_{u \in V_1} \Delta_u$.
Then we have through standard sampling arguments that $\mathbb E[\Delta] \leq \mathcal O(\e) |\delta(V_1)|$.
Thus, in spirit, optimizing $\sigma_1$ against our estimates $\{h_u\}_{u \in V_1}$ is not so different from optimizing against the optimal $\{h_u'\}_{u \in V_1}$. 
After computing $\sigma_1$ recursively, we can then assign unary rewards according to $\sigma_1$ to $V_2$, and recurse on $V_2$ to obtain $\sigma_2$, which we combine with $\sigma_1$ to produce our final labeling $\sigma : V \to \zo$. 
Note that each edge of $E$ crosses exactly one of the cuts considered in this algorithm, and so our total expected error is at most $\mathcal O(\e) |E|$, i.e., $\mathbb E[\val{\sigma}] \geq \opt - \mathcal O(\e) |E|$.
The multiplicative approximation guarantee then follows from noting $\opt \geq \frac{|E|}{4}$, as a random labeling satisfies each constraint with probability at least $\frac{1}{4}$.

\paragraph{Bounding runtime.}
A careful reader may have noticed that we have yet to use the fact that $\rho(V_1) \leq r$. 
Indeed, this fact is used only to bound the runtime.
The crux of this is designing a sampling strategy that results in a bounded size of the sample set $S$.
As a warm-up, notice that the collection of neighborhoods
\[
    \{ N(u) \cap V_2 \mid u \in V_1\}
\]
has size at most $2^r$, as each neighborhood corresponds to some row in the cut matrix $A_{V_1}$, and the span of these rows contains at most $2^r$ vectors. 
Thus, if we ``share'' sampled sets $S_u = S_{u'}$ for vertices $u, u' \in V_1$ with $N(u) \cap V_2 = N(u') \cap V_2$, we can bound $|S| \leq 2^r \e^{-2}$. 
Let $T(n)$ be the run-time of our algorithm on inputs with $n$ vertices. 
We can then bound
\[
    T(n) \leq 2^{2^r \e^{-2}} (2 T(2n/3)) + \operatorname{poly}(n),
\]
yielding runtime $n^{\mathcal O(2^r \e^{-2})}$. 
To achieve the improved bound of $n^{\mathcal O(r \e^{-2})}$, we need to bound $|S| \leq r \e^{-2}$, which requires a more careful sampling strategy. 
We begin by randomly permuting $V_2$.
Now, for each $u \in V_1$, let its sampled set $S_u$ be the first $ \e^{-2}$ members of $N(u) \cap V_2$ in this sorted order, and $S = \cup_{u \in V_1} S_u$ as usual.

We claim the bound $|S| \leq r \e^{-2}$.
Recall that every vertex $v \in V_2$ corresponds to a column $C_v$ of the cut matrix $A_{V_1}$. 
We will now partition the (multi)set of columns $\set{C_v}_{v \in S}$ into $\e^{-2}$ independent (multi)sets --- then, as $A_{V_1}$ has rank at most $r$, each of these must contain at most $r$ column vectors, implying the desired bound $|S| \leq r \e^{-2}$.
We will iteratively place the columns corresponding to vertices in $S$ according to the selected ordering of $V_2$.
Suppose we are currently considering column $C_v$ for $v \in S$.
Then $v$ was sampled because $v \in S_u$ --- $v$ is one of the first $\e^{-2}$ neighbors of $u$ in $V_2$.
Thus, at least one of our independent sets does not contain a column corresponding to a neighbor of $u$, and $C_v$ cannot be in the span of that independent set, so add $C_v$ to it.

\subsection{Interval Graphs are Nearly Bounded Rank-Width}

In this subsection, we give an overview of the techniques used to prove the following structural theorem about interval graphs.

\begin{structuralrestatement}
    \intervalstructurestatement
\end{structuralrestatement}

\paragraph{Warmup: $\mathcal O(\e^{-1} \log n)$ rank-width.}
We begin with a simple warmup by showing how to delete at most $\e |E|$ edges and obtain a graph of rank-width $\mathcal O(\e^{-1} \log n)$.
Throughout, we will suppose that $\e^{-1}$ is a power of two.
Given an interval representation of $G$, order the vertices $V$ by non-decreasing order of their left endpoints.
We identify a vertex $v \in V$ with its position in the ordering.
Thus, we can denote the vertex immediately after $v$ in the ordering as $v + 1$, and say $u < v$ if $u$ precedes $v$ in the ordering.
For a vertex $v \in V$, denote the \EMPH{forward neighborhood} of $v$ as
\[
    N^+(v) := \{u \in V \mid vu \in E, v < u\}.
\]
Letting $d^+_v := |N^+(v)|$ be the \EMPH{forward degree} of $v$, we can write $N^+(v) = [v + 1, v + d^+_v]$.
Our strategy is now to normalize the forward endpoint of each vertex.
Define the \EMPH{scale} of $v$ to be $s_v$, the power of two such that $s_v \leq d_v^+ < 2 s_v$, or $0$ if $d^+_v = 0$.
Now, for each vertex $v \in V$ with $s_v \geq \e^{-1}$, decrease $d^+_v$ by deleting edges until $v + d^+_v$ is a multiple of $\e s_v$.
We have that the number of deleted edges from $N^+(v)$ is at most $\e s_v \leq \e d^+_v$, so the total number of deleted edges is at most $\e |E|$.

We claim that the resulting graph has rank-width at most $\mathcal O(\e^{-1} \log n)$.
To show this, we produce a rank-decomposition and bound its width.
Let $T$ be a caterpillar with $n$ leaves. 
Assign vertices in $V$ to the leaves of $T$ according to our computed ordering of $V$. %
There are two types of cuts in this rank-decomposition.
For any vertex $v \in V$, there is the cut $\{v\}$, which has rank $\rho(\{v\}) \leq 1$.
Additionally, for any $i \in V$, there is the cut $S_i := \{i+1, i+2, \ldots, n\}$. 
Let us now bound $\rho(S_i)$.
Consider any $v \not\in S_i$ with scale $s_v$.
If $N(v) \cap S_i \neq \varnothing$, then there are at most $2\e^{-1}$ possible endpoints of $v$'s forward neighborhood $[v+1, v + d^+_v]$ as there are at most $2 \e^{-1}$ multiples of $\max\{\e s_v, 1\}$ within distance $d^+_v < 2 s_v$ to the right of position $i$.
Thus, over all vertices $v \not \in S_i$ with scale $s_v$, there are at most $2 \e^{-1}$ unique non-empty neighborhoods $N(v) \cap S_i$.
The rank of $S_i$ is upper-bounded by the number of these unique neighborhoods, as these are the number of distinct non-zero columns in the matrix $A_{S_i}$.
Summing over all $\log_2(n)$ scales yields the bound $\rho(S_i) \leq \mathcal O(\e^{-1} \log n)$. 

\begin{figure}[h]
    \centering
    \begin{tikzpicture}[
        font=\small,
        vertex/.style={circle, draw, fill=white,
            minimum size=5.5mm, inner sep=0pt, line width=0.5pt},
        edge/.style={line width=0.5pt}
    ]
        \node[vertex] (a) at (1.4,1) {$a$};
        \node[vertex] (b) at (0,0) {$b$};
        \node[vertex] (c) at (-1.2,-1.55) {$c$};
        \node[vertex] (d) at (0.1,-1.55) {$d$};
        \node[vertex] (e) at (1.4,-1.55) {$e$};
        \foreach \u/\v in {a/b,b/c,b/d,b/e}
            \draw[edge] (\u) -- (\v);

        \draw[->, line width=0.8pt] (2.25,-0.3) -- (3.9,-0.3);

        \begin{scope}[xshift=6.1cm]
            \node[vertex] (ar) at (1.4,1) {$a$};
            \node[vertex] (b1) at (-0.25,0.25) {$b_1$};
            \node[vertex] (b2) at (0.55,-0.45) {$b_2$};
            \node[vertex] (cr) at (-1.2,-1.55) {$c$};
            \node[vertex] (dr) at (0.1,-1.55) {$d$};
            \node[vertex] (er) at (1.4,-1.55) {$e$};
            \foreach \u/\v in {ar/b1,b1/cr,b2/dr,b2/er}
                \draw[edge] (\u) -- (\v);
            \draw[line width=0.9pt] (b1) -- (b2);
        \end{scope}
    \end{tikzpicture}
    \caption{Splitting a degree-four node $b$ into two degree-three nodes $b_1$ and $b_2$.}
    \label{fig:vertex-split}
\end{figure}
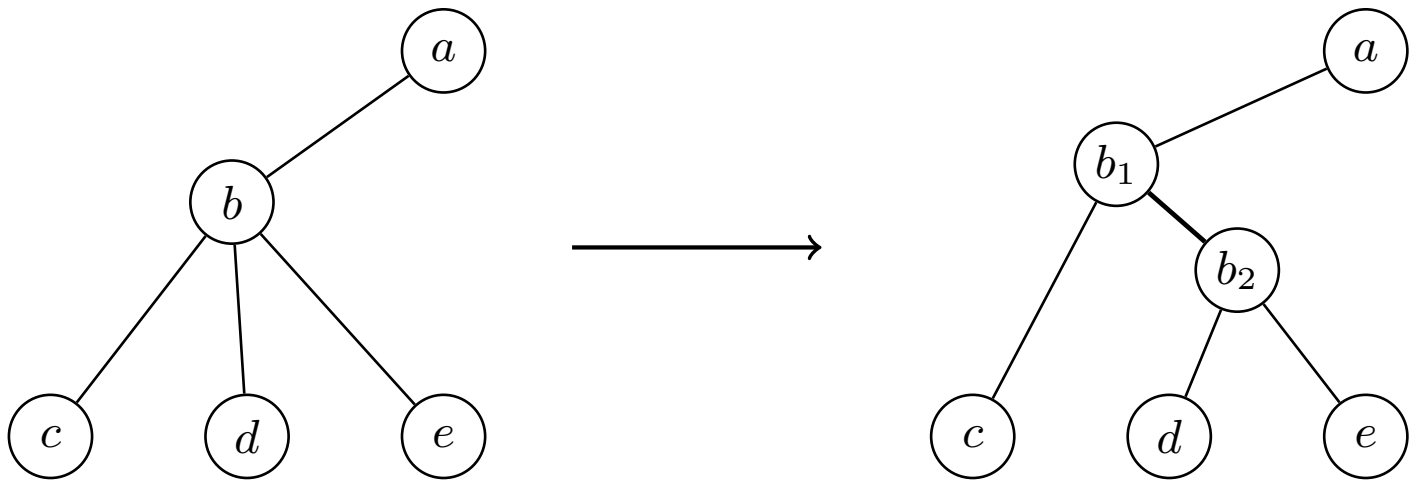

\paragraph{A different type of rank-decomposition.}  
The main difficulty in reducing the rank-width from $\mathcal O(\log n)$ to $\mathcal O(1)$ is that different scales do not play nicely together. 
To deal with this, we will need a more interesting tree for our rank-decomposition than a caterpillar. 
Specifying an exact rank-decomposition is often quite tedious, so we introduce a more general form of rank-decomposition, which can be transformed into a standard rank-decomposition with similar bounds.
We again begin with a sub-cubic tree $T$, but we can now assign vertices of $V$ to \emph{any} node of $T$, with the following restrictions:
\begin{itemize}
    \item Each node of $T$ should have at most $Q$ vertices of $V$ assigned to it.
    \item Each edge $e \in E(T)$'s implied cut $S_e$ should have rank $\rho_G(S_e) \leq r$.
\end{itemize}
We claim that, from such a decomposition, we can obtain a rank-decomposition of $G$ with width at most $r + Q$.
To do so, we add any subtree with $Q$ leaves to each vertex of $T$, and match these leaves arbitrarily with the vertices of $V$ assigned to that node of $T$.
Note that this may increase the degree of some nodes in $T$ to $4$ --- to reduce degrees back down to $3$, we split any such degree-four node into two new nodes, see \Cref{fig:vertex-split}.
Call this new tree $T'$ and consider any implied cut $S_f$ for $f \in E(T')$. 

There are three types of cuts this can correspond to.
In the first case, $S_f$ corresponds to a cut $S_e$ with $e \in E(T)$ and so $\rho_G(S_f) = \rho_G(S_e) \leq r$
In the second case, $f$ is an edge internal to some extra subtree we added to $T$, in which case $|S_e| \leq Q$ and so $\rho_G(S_e) \leq Q$.
Finally, if $f$ is an edge between two copies of a node we produced in the degree-reduction step, we have that $S_f$ is the same as the cut on the original parent edge, except that the vertices assigned to the node have moved to the other side of the cut. 
Moving one vertex across a cut changes the rank by at most one, so $\rho(S_f) \leq r + Q$.

\begin{figure}[h]
    \centering
    \includegraphics[width=0.5\linewidth]{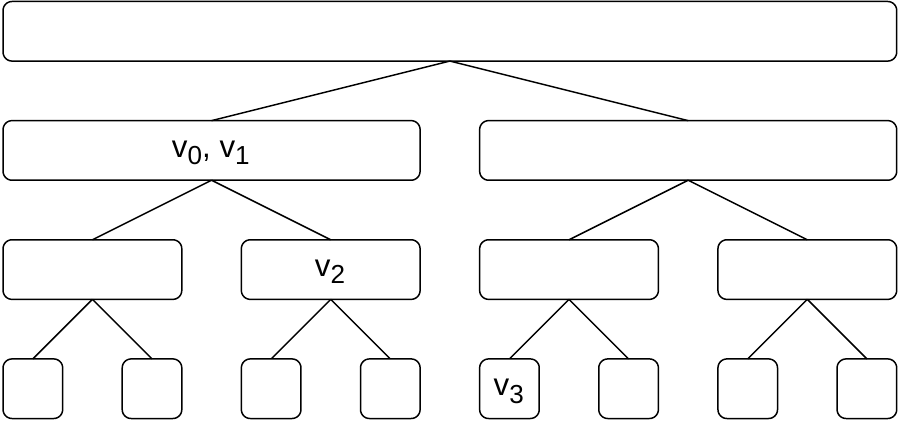}
    \caption{An excerpt of $T$ with $\e = \frac{1}{2}$ and $\theta = 1$ when $G = K_4$, the complete graph on $4$ vertices. Here, $s_{v_0} = s_{v_1} = 2, s_{v_2} = 1$, and $s_{v_3} = 0$.}
    \label{fig:segtree}
\end{figure}

\paragraph{Our rank-decomposition.}
We now construct our rank-decomposition.
After this construction, we will show how to delete edges to bound the resulting width.
For simplicity, we assume that $n$ is a power of two. 
Let $T$ be the complete rooted binary tree with $2 \e^{-1} n$ leaves.
We now sample a uniformly random integer offset $\theta \in [0, n \e^{-1}) \cap \mathbb N$, and temporarily assign the $i$th vertex in our ordering to the $(i + \theta)$th leaf of $T$. 
For $v$ with scale $s_v \neq 0$, we then move $v$ up $\log_2(\e^{-1} s_v)$ levels of the tree from its leaf.  
See \Cref{fig:segtree} for an example.
The tree $T$ should be interpreted as a ``segment tree'', where each node corresponds to some sub-interval of $[0, 2n-1]$. 
We refer to these nodes as \EMPH{cells}.
We call such a cell an $\alpha$-cell if it contains exactly $\alpha$ leaves.
Then we have that $v$ is assigned to the unique $(\e^{-1} s_v)$-cell $[k \cdot (\e^{-1} s_v), (k+1) \cdot (\e^{-1} s_v) - 1]$ containing $v + \theta$.
Call this the \EMPH{home cell}, $C_v$, of $v$.
Note that this decomposition currently violates both our properties --- there can exist nodes of $T$ with many vertices of $V$ assigned to them, and cuts $S_e$ may have high rank.
We will now show how edges may be deleted to fix these problems.

\paragraph{Restricting forward neighborhoods to home cells.}
We aim to limit the behavior of vertices within a home cell, with an eventual eye to showing each home cell can contain a bounded number of vertices after removing twins.
To this end, for each $v \in V$ with home cell $C_v$, we delete all edges in $v$'s forward neighborhood which go to vertices $u \not \in C_v$. 
Since the random shift $\theta$ is uniformly random modulo $|C_v|$, the position of $v$ within its home cell is uniformly random.
For any forward neighbor $u \in N^+(v)$, at most $u -v \leq d^+_v$ of these positions place $u$ outside of $C_v$.
So we have that each forward edge is deleted with probability at most $d_v^+ / |C_v| \leq 2 \e$.
Thus, by linearity of expectation, this step deletes at most $2 \e |E|$ edges in expectation.
Choosing a $\theta$ with minimum loss then yields a deterministic loss of at most $2 \e |E|$.

\paragraph{Normalizing forward neighborhoods within a home cell.}
Even after restricting forward neighborhoods within home cells, there still remain $\mathcal O(|C_v|^2)$ possible forward neighborhoods within a home cell $C_v$. 
We now describe deletions which reduce the possibilities to $\e^{-4}$.
For a vertex $v \in V$ with scale $s_v \geq \e^{-1}$, we aim to coarsify $N^+(v)$ to be a union of $\e s_v$-cells. 
To this end, we delete any edge in $v$'s forward neighborhood to some $u \in N^+(v)$ which shares an $\e s_v$-cell with either $v$ or $v + d_v^+ + 1$. %
This process deletes at most $2 \e s_v \leq 2 \e d_v^+$ edges from the forward neighborhood of $v$, so the total number of edges deleted is at most $2 \e |E|$.
After this step, the forward neighborhood of $v$ is now a range of $\e s_v$-cells. 
Because $C_v$ contains exactly $\e^{-2}$ cells of size $\e s_v$, this implies there are at most $\mathcal O(\e^{-4})$ possible forward neighborhoods for vertices with home cell $C_v$.

\paragraph{What about backwards neighborhoods?}
We cannot immediately conclude that each home cell contains $\mathcal O(\e^{-4})$ ``types'' of vertices (i.e., vertices with different neighborhoods), as there could be a large variety of ``backwards'' neighborhoods of these vertices, despite their bounded possibilities for forward neighborhoods.
To deal with this, we introduce a third and final edge deletion step.
For a vertex $u \in V$, delete the edge $uv$ for $v \in N^+(u)$ if $s_u \leq \e s_v$. 
After this step, every edge in $v$'s backwards neighborhood must have comparable scale to $v$, which we will see is sufficient to conclude our bounded rank-width argument.
For $v \in V$, let us count the number of deleted edges in the backwards neighborhood of $v$ from this step.
Because $s_u \leq \e s_v$, we have that $d^+_u < 2 s_u \leq 2 \e s_v \leq 2 \e d^+_v$.
Thus, the only vertices $u$ which could lead to deleted edges $uv$ must be within $2 \e d^+_v$ positions to the left of $v$.
Because there are at most $2 \e d^+_v$ of these vertices, we delete at most $2 \e |E|$ edges in total.

\paragraph{Bounding the size of home cells.}
Suppose that there are two vertices $u, v \in V$ with the same neighborhood (i.e., they are false twins).
Then notice that deleting one of $u$ and $v$ will not affect the rank-width of the graph.
Thus, we delete vertices until there are no false twins remaining in our graph before bounding the rank-width.

Consider any vertex $v$ with home cell $C_v$.
We now wish to bound the number of vertices which are assigned to this home cell $C_v$.
Let $\ell_v$ be the $\max\{1, \e^2 s_v\}$-cell containing $v$, and $r_v$ be the final $\max\{1, \e s_v\}$-cell of $v$'s forward neighborhood, after our normalizations of forward neighborhoods. 
Suppose that $u$ has the same home cell $C_u = C_v$, and additionally has the same cells $\ell_u = \ell_v$ and $r_u = r_v$.
Note also that $s_u = s_v$.
Then we claim that $u$ and $v$ are false twins. 
First, we note that $u$ and $v$ are not adjacent because they share the same $\e s_v$-cell. 

Now consider any other $w \in V$. 
We will show that either $w$ is adjacent to both of $u$ and $v$, or neither of them.
Suppose without loss of generality that $u < v$.
In the first case, we have that $u < v < w$.
Then $w$ is adjacent to $v$ if and only if $s_v > \e s_w$ and $w$'s $\e s_v$-cell lies in the range $(\ell_v, r_v]$, which are exactly the same conditions for $w$ being adjacent to $u$.
In the second case, $u < w < v$, we have that $w$ is in the same $\e s_v$-cell as $u$, and so was pruned from $u$'s forward neighborhood.
If $s_w \leq \e s_v$, then $w$ was pruned from the backwards neighborhood of $v$. 
Otherwise, if $s_w > \e s_v$, then $\e s_w > \e^2 s_v$. Since $u$ and $v$ share the same $\e^2 s_v$-cell, we have that $w$ and $v$ share the same $\e s_w$-cell, and thus $v$ was pruned from the forward neighborhood of $w$.
Thus, $w$ is adjacent to neither of $u$ or $v$ in this case.
Finally, in the case $w < u < v$, we have that $u$ and $v$ share the same $\e s_w$-cell, and so should either both or neither be adjacent to $w$.

Note that the number of choices for $\ell_v$ is $\e^{-3}$ and the number of choices for $r_v$ is $\e^{-2}$, so this argument implies there are at most $\mathcal O(\e^{-5})$ vertices with home cell $C_v$ after removing false twins.

\paragraph{Bounding the rank of cuts.}
Consider the implied cut $S_e$ for an edge $e \in E(T)$ of our tree decomposition $T$.
We wish to bound the rank $\rho(S_e)$. 
Let $C$ be the cell that is the child endpoint of $e$.
Then, letting $V_C := \set{v \in V \mid C_v \subseteq C}$ be the vertices with home cell contained in $C$'s subtree, we have that $V_C = S_e$.
To bound the rank $\rho(V_C)$, we bound the number of different neighborhoods that $u \in V \setminus V_C$ can have within $C$.
To this end, fix some $u \in V \setminus V_C$ and let $C_u$ be the home cell of $u$.
If $C_u \cap C = \varnothing$, then $u$ is not adjacent to any of $C$ due to our edge deletions.
Note that the cells form a laminar family of sets, so the only remaining case is $C \subseteq C_u$. 

We first consider the case where $|C_u| \geq \e^{-2} |C|$.
Then we have that $s_u = \e |C_u| \geq \e^{-1} |C|$.
We will show that, in this case, either $u$ is adjacent to all of $V_C$ or none of $V_C$.
If $u < v$ for all $v \in V_C$, then all neighbors of $u$ in $V_C$ are part of $u$'s forward neighborhood.
Recall that $u$'s forward neighborhood is composed of whole $\e s_u \geq |C|$-cells, so either all of $C$ is in this neighborhood or none of it.
Note that if $v < u$ for some $v \in V_C$, then $v$ is not adjacent to $u$ because $s_v \leq \e |C| \leq \e s_u$, and so the edge would have been pruned in our final deletion step.
Thus, if $u > v$ for all $v \in V_C$, then $u$ has no edges to $C$.
Finally, suppose $u$ lies somewhere in the middle of $V_C$. 
Then we have that $u \in C$.
Then, for any $v \in V_C$ with $u < v$, we have that $u$ and $v$ share the same $\e s_u \geq |C|$-cell, so the edge $uv$ would have been deleted. 
Thus, in this case as well, $u$ has no edges to $V_C$.

Now consider the case where $|C_u| < \e^{-2} |C|$.
Here we can no longer say anything interesting about the neighborhood of $u$ in $V_C$.
However, note that there are only $\mathcal O(\log \e^{-1})$ such cells $C_u$, each of which contain at most $\mathcal O(\e^{-5})$ vertices.
Thus, we can bound $\rho(V_C) \leq \mathcal O(\e^{-5} \log \e^{-1})$.
So we have shown that the rank-width of the graph after our $6 \e |E|$ edge deletions is at most $\mathcal O(\e^{-5} \log \e^{-1})$.

\section{A PTAS for Bounded Rank-Width Graphs}
In this section, we give a PTAS for \csp on bounded rank-width graphs.

\begin{ptasrestatement}
    \rankwidthptasstatement
\end{ptasrestatement}

More precisely, we solve the following generalization of Boolean Max-2-CSP.
In addition to the input graph $G = (V, E)$ and edge rewards $\{f_{uv}\}_{uv \in E}$, we are given non-negative unary rewards $\{g_v\}_{v \in V}$, where $g_v : \{0, 1\} \rightarrow \mathbb R_{\geq 0}$.
The goal of this modified problem is to find  a labeling $\sigma : V \rightarrow \{0, 1\}$ which maximizes
\[
    \val{\sigma} = \sum_{uv \in E} f_{uv}(\sigma(u), \sigma(v)) + \sum_{v \in V} g_v(\sigma(v)).
\]

We prove the following generalization of \Cref{thm:ptas}.

\begin{theorem} \label{thm:unaryptas}
    Given an instance of \csp with constraint graph $G = (V, E)$, a rank-decomposition of $G$ with width at most $r$, binary rewards $\{f_{uv}\}_{uv \in E}$, unary rewards $\set{g_v}_{v \in V}$, and $\e > 0$, there is a randomized $n^{\mathcal O(r \e^{-2})}$-time algorithm that computes a labeling $\sigma : V \to \zo$ with \[\mathbb E[\val{\sigma}] \geq \opt - \e |E|.\]
\end{theorem}

Note that \Cref{thm:ptas} follows directly from \Cref{thm:unaryptas} by setting $g_v = 0$ for all $v \in V$ and observing $\opt \geq \frac{|E|}{4}$.

\subsection{The Algorithm} \label{subsection:algorithm}
In this subsection, we present our algorithm for \csp.

\paragraph{Base case and cuts.} If $|V| = 1$, then there are no edges in $G$, so we simply set \[
    \sigma(v) := \operatorname{argmax}_{b \in \{0, 1\}} g_v(b).
\]
Otherwise, we have that $|V| \geq 2$, and so can obtain a partition $V_1, V_2 \subseteq V$ from our rank-decomposition with properties
\[
    |V_1|, |V_2| \geq \frac{n}{3} \qquad \text{and} \qquad \rho(V_1) \leq r.
\]

\paragraph{Choosing a sample set $S$.}
Our algorithm now sets $q := \lceil 4 \e^{-2} \rceil$ and samples a uniformly random ordering of $V_2$. 
For each $u \in V_1$, we set $S_u \subseteq V_2$ to be the first $\min \{q, |N(u) \cap V_2|\}$ neighbors of $u$ in $V_2$ according to our sampled ordering.
We then set $S := \cup_{u \in V_1} S_u$.

\begin{observation} \label{obs:uniformsampling}
    For each $u \in V_1$, we have that $S_u$ is a uniformly sampled size-$|S_u|$ subset of $N(u) \cap V_2$.
\end{observation}

\begin{lemma} \label{lem:boundsizeofs}
    For any ordering of $V_2$, we have that $|S| \leq r q$. 
\end{lemma}
\begin{proof}
    Let $A$ be the adjacency matrix of the cut defined by $V_1$, with rows corresponding to vertices $u \in V_1$ and columns corresponding to vertices $v \in V_2$. 
    For $v \in V_2$, let $C_v$ be the column associated with $v$. 
    To show that $|S| \leq r q$, we will partition the multi-set of vectors $\{C_v\}_{v \in S}$ into $q$ independent sets --- because $A$ has rank at most $r$, each of these independent sets must have size at most $r$, for a total size of at most $rq$.

    We will build these independent sets iteratively, by processing $S$ in the sampled order of $V_2$.
    Suppose we are considering $v \in S$. 
    We have that $v$ was sampled because it is one of the first $q$ neighbors in $V_2$ of some $u \in V_1$.
    Thus, there are at most $q-1$ previously processed vertices adjacent to $u$ --- so there is at least one independent set without any vector that has a $1$ in the $u$th position.
    This implies that $C_v$ is not in the span of that independent set, and we can add $C_v$ without breaking independence.
    Thus, we can iteratively add $C_v$ for all $v \in S$ to some independent set without breaking independence, yielding the desired bound $|S| \leq r q$.
\end{proof}

\paragraph{Guessing a labeling for $S$.} 
Our algorithm now iterates over all labelings $\sigma_S : S \to \zo$ of $S$, and for each performs the following steps:
\begin{enumerate}
\item For each $u \in V_1$ and $b \in \{0, 1\}$, define
        \[
            h_u(b) := \frac{|N(u) \cap V_2|}{|S_u|} \sum_{v \in S_u} f_{uv}(b, \sigma_S(v)),
        \]
        or $h_u(b) = 0$ if $S_u = \varnothing$.
        Recurse on $G[V_1]$ with unary rewards $g'_u(b) := g_u(b) + h_u(b)$, and let $\sigma_1 : V_1 \rightarrow \{0, 1\}$ be the computed labeling.

        \item For each $v \in V_2$ and $b \in \{0, 1\}$, define
        \[
            h_v(b) := \sum_{u \in N(v) \cap V_1} f_{uv}(\sigma_1(u), b).
        \]
        Recurse on $G[V_2]$ with unary rewards $g'_v(b) := g_v(b) + h_v(b)$, and let $\sigma_2 : V_2 \rightarrow \{0, 1\}$ be the computed labeling.

        \item Merge $\sigma_1$ and $\sigma_2$ to create a labeling $\sigma : V \to \zo$.
\end{enumerate}

Our algorithm then returns the labeling $\sigma$ with highest value among all $2^{|S|}$ computed labelings.
We refer to our algorithm as $\mathcal{A}$, and let $\mathcal{A}(G)$ refer to $\val{\sigma}$ for the returned labeling $\sigma$.

\paragraph{Running time.}
We now show that the running time of our algorithm is polynomial for fixed $r$ and $\e$.
\begin{lemma}
    The running time of $\mathcal A$ is $n^{\mathcal O(r \e^{-2})}$.
\end{lemma}
\begin{proof}
    Let $T(n)$ indicate the maximum runtime of $\mathcal A$ on inputs with $n$ vertices. 
    Then $T(1) = \mathcal O(1)$ and 
    \[
        T(n) \leq 2^{|S|} \left(2 T\left(\frac{2n}{3}\right)\right) + \operatorname{poly}(n) \leq 2^{\mathcal O(r \e^{-2})} T\left(\frac{2n}{3}\right) + \operatorname{poly}(n)
    \]
    by \Cref{lem:boundsizeofs}.
    Solving yields $T(n) \leq 2^{\mathcal O(\log n \cdot r \e^{-2})} \leq  n^{\mathcal O(r \e^{-2})}$ as wanted.
\end{proof} 

\subsection{Analysis of Loss}
We now aim to show that $\mathbb E[\mathcal A] \geq \opt - \e |E|$ by induction on $|V|$.
When $|V| = 1$, the only rewards are unary, which we optimize exactly. 
We now consider the $|V| \geq 2$ case.
To aid in bounding the performance of $\mathcal A$, let us fix an optimal labeling $\sigma_{\opt} : V \to \zo$.

\paragraph{Estimating the cut rewards.}
Consider the branch of $\mathcal A$ in which $\sigma_S$ agrees with $\sigma_{\opt}$ on $S$. 
For $u \in V_1$ and $b \in \{0, 1\}$, let $h_u(b)$ and $g'_u(b)$ be defined as in that branch. 
Additionally, we define
\[
    h_u'(b) := \sum_{v \in N(u) \cap V_2} f_{uv}(b, \sigma_{\opt}(v))
\]
as the cut rewards when $\sigma_{\opt}$ is the labeling used for $V_2$. 
We aim to show that $h_u(b)$ and $h_u'(b)$ are not so different.
To this end, we define the maximum error at $u$ as
\[
    \Delta_u := \max_{b \in \zo} |h_u(b) - h_u'(b)|
\]
and the total error as $\Delta := \sum_{u \in V_1} \Delta_u$.

\begin{lemma} \label{lem:newbounddelta}
    We have that $\mathbb E[\Delta] \leq |\delta(V_1)| / \sqrt{q}$.
\end{lemma}
\begin{proof}
    Fix $u \in V_1$ and let $d_u = |N(u) \cap V_2|$.
    If $d_u \leq q$, then $S_u = N(u) \cap V_2$ and so $\Delta_u = 0$. 

    Suppose $d_u > q$ and fix some label $b \in \zo$. 
    Consider the random variable $h_u(b)/d_u$. 
    Observation~\ref{obs:uniformsampling} says that $ h_u(b) / d_u$ is the average of $q$ edge rewards uniformly sampled without replacement. 
    Thus, this random variable $h_u(b) / d_u$ has mean $h_u'(b) / d_u$ and variance at most $\frac{1}{4q}$. 
    So we have that
    \[
        \mathbb E[|h_u(b) - h_u'(b)|] \leq \frac{d_u}{2 \sqrt q}
    \]
    and thus
    \[
        \mathbb E[\Delta_u] \leq \mathbb E[|h_u(0) - h_u'(0)|] + \mathbb E[|h_u(1) - h_u'(1)|] \leq \frac{d_u}{\sqrt q}.    \]
    Summing over all $u \in V_1$ and noting $\sum_{u \in V_1} d_u = |\delta(V_1)|$ completes the proof.
\end{proof}

\paragraph{Bounding loss from combining $\sigma_1$ and $\sigma_2$.}
We again consider only the branch where $\sigma_S$ agrees with $\sigma_{\opt}$ on $S$.
Let $\opt_1$ and $\opt_2$ denote the optimal values of our produced instances for $G[V_1]$ and $G[V_2]$ in this branch.
For $i \in \{1, 2\}$ and any labeling $\sigma : V_i \to \zo$, let $\operatorname{Value}_i(\sigma)$ be the value of $\sigma$ on the produced instance for $G[V_i]$.
Now let
\[
    L_i := \opt_i - \operatorname{Value}_i(\sigma_i)
\]
be the loss of $\mathcal A$ on instance $G[V_i]$. 
We have by induction that $\mathbb E[L_i] \leq \e |E[V_i]|$, which we will use later when bounding $\mathbb E[\mathcal A]$.

\begin{lemma} \label{lem:boundsigma}
    Let $\sigma$ be the labeling obtained by merging $\sigma_1$ and $\sigma_2$.
    We have that \[\val{\sigma} \geq \opt - L_1 - L_2 - 2 \Delta.\]
\end{lemma}

\begin{proof}
    
    Because the instance $G[V_2]$ counts the exact cut rewards from $\delta(V_1)$, we have that
    \[
        \val{\sigma} = \operatorname{Value}_1(\sigma_1) + \operatorname{Value}_2(\sigma_2) - \sum_{u \in V_1} h_u(\sigma_1(u)).
    \]
    To get a handle on the $h_u$ terms, we will compare them to $h_u'$. 
    To this end, we define, for any $\tau : V_1 \to \zo$, the difference function
    \[
        D(\tau) := \sum_{u \in V_1} \left( h_u(\tau(u)) - h_u'(\tau(u)) \right).
    \]
    Note that we have $|D(\tau)| \leq \Delta$ for any $\tau$. 
    Now we can express 
    \[
        \val{\sigma} = \operatorname{Value}_1(\sigma_1) + \operatorname{Value}_2(\sigma_2) - \sum_{u \in V_1} h_u'(\sigma_1(u)) - D(\sigma_1).
    \]
    Recall that for $i \in \{1, 2\}$,
    \[
        \operatorname{Value}_i(\sigma_i) = OPT_i - L_i \geq \operatorname{Value}_i(\sigma_{\opt}) - L_i,
    \]
    where $\operatorname{Value}_i(\sigma_{\opt})$ evaluates the restriction of $\sigma_{\opt}$ to $V_i$.
    Thus, we have that
    \[
        \val{\sigma} \geq \operatorname{Value}_1(\sigma_{\opt}) + \operatorname{Value}_2(\sigma_{\opt}) - L_1 - L_2 - \sum_{u \in V_1} h_u'(\sigma_1(u)) - D(\sigma_1).
    \]
    Also recall that the rewards of $G[V_1]$ and $G[V_2]$ both include rewards from the cut $\delta(V_1)$ --- $G[V_1]$ estimates these rewards with error corresponding to $D$, while $G[V_2]$ computes it precisely according to $\sigma_1$.
    Thus, we can write
    \[
        \operatorname{Value}_1(\sigma_{\opt}) + \operatorname{Value}_2(\sigma_{\opt}) = \opt + D(\sigma_{\opt}) + \sum_{u \in V_1} h_u'(\sigma_{1}(u)).
    \]
    Substituting into our previous bound for $\val{\sigma}$ yields
    \[
        \val{\sigma} \geq \opt - L_1 - L_2 + D(\sigma_{\opt}) - D(\sigma_1) \geq \opt - L_1 - L_2 - 2 \Delta.
    \]
\end{proof}

\begin{corollary}
    Let $\sigma$ be the labeling obtained by merging $\sigma_1$ and $\sigma_2$.
    We have that
    \[
        \mathbb E[\val{\sigma}] \geq OPT - \e |E|.
    \]
\end{corollary} 

\begin{proof}
    Apply expectation to the bound given by \Cref{lem:boundsigma}
    \[
        \mathbb E[\val{\sigma}] \geq \opt - \mathbb E[L_1] - \mathbb E[L_2] - 2 \mathbb E[\Delta].
    \]
    Applying induction and \Cref{lem:newbounddelta} yields the bound
    \[
        \mathbb E[\val{\sigma}] \geq \opt - \e |E[V_1]| - \e |E[V_2]| - 2 |\delta(V_1)| / \sqrt{q} \geq \opt - \e |E|.
    \]
\end{proof}

Because $\mathbb A$ returns the best computed labeling over all branches, its final output will be at least as good as $\sigma$.
Thus, $\mathbb E[\mathcal A] \geq \opt - \e |E|$, as wanted.

\subsection{Extension to Bounded-Alphabet CSPs}

Our algorithm can be modified to support CSPs over a bounded alphabet size $k$. 
We will briefly show that one can obtain a $(1 - \e)$-approximation with running time $n^{\mathcal O(r \e^{-2} k^6 \log k)}$.
Let $[k] = \{0, 1, \ldots, k-1\}$ be the alphabet over which the CSP is defined.
As before, we suppose that each edge $uv \in E$ has a non-zero constraint $f_{uv} : [k]^2 \to \zo$.
We again sample $S$ as in \Cref{subsection:algorithm} and iterate over all $k^{|S|}$ labelings of $S$. 
For each guess, we define the unary rewards $h_u(b)$ as before, but now for every $b \in [k]$.
The bound on $\Delta$ proceeds as in \Cref{lem:newbounddelta}, but we now have
\[
    \mathbb E[\Delta] \leq \frac{k |\delta(V_1)|}{2 \sqrt q}
\]
because we must sum error over all $k$ labels. 
Thus, we must set $q \approx k^2 \e^{-2}$ to keep error below $\e |E|$. 
The rest of the proof proceeds as follows, except the conversion from $\e |E|$ to a $(1 - \e)$-approximation.
Previously, we argued these were equivalent up to constant factors because $\opt \geq \frac{|E|}{4}$, as each constraint is satisfied with probability at least $\frac{1}{4}$ by a random labeling.
The same argument now only yields $\opt \geq \frac{|E|}{k^2}$. 
Thus, to keep error below $\e |E|$, we must set $q \approx k^6 \e^{-2}$. 
Let $T(n)$ be the maximum running time on $n$ vertices as before.
We can then express the recurrence
\[
    T(n) \leq k^{rq} \left( 2T \left( \frac{2n}{3} \right)  + \operatorname{poly}(n) \right).
\]
Solving yields a running time bound of of
\[
    n^{\mathcal O(rq \log k)} = n^{\mathcal O(r \e^{-2} k^6 \log k)}.
\]

\section{Interval Graphs are Almost Bounded Rank-Width}

In this section, we show that interval graphs are only a few edge deletions away from having bounded rank-width.
\begin{structuralrestatement}
    \intervalstructurestatement
\end{structuralrestatement}

\subsection{Deleting Edges}

    \paragraph{Forward neighborhoods and scales.}
    Without loss of generality, suppose that $\e^{-1}$ is a power of two. 
    This can be done by re-scaling, as the resulting change in rank-width is only a constant factor.
    Given an interval representation of $G$, order the vertices $V$ by non-decreasing order of their left endpoints.
    We identify a vertex $v \in V$ with its position in this ordering and write $u < v$ when $u$ appears before $v$ in the ordering.
    In particular, we can denote the vertex immediately after $v$ in this ordering as $v+1$, the next vertex as $v+2$, and so on.
    For a vertex $v \in V$, denote the \EMPH{forward neighborhood} of $v$ as
    \[
        N^+(v) := \{u \in V \mid vu \in E, v < u\}.
    \]
    Similarly, denote the \EMPH{forward degree} of $v$ as $d^+_v := |N^+(v)|$. 
    The advantage of this ordering of $V$ is that the forward neighborhood of any vertex is simple to describe;
    we have that $N^+(v) = [v+1, v + d^+_v] \cap \mathbb N$.
    We will often abuse notation and simply write $N^+(v) = [v + 1, v + d^+_v]$, leaving the intersection with $\mathbb N$ implicit.
    We denote the \EMPH{scale} of $v$ as the largest power of two at most $d^+_v$. 
    Formally, define the scale of $v$ as
    \[
        s_v := \max \{ 2^i \mid i \in \mathbb N, 2^i \leq d^+_v\},
    \]
    or $s_v = 0$ if $d^+_v = 0$.
    In particular, we then have that $s_v \leq d^+_v < 2 s_v$ for non-zero $s_v$.
    Roughly speaking, we will aim to normalize based on scale: if $u$ and $v$ are ``close'' together in the ordering and $s_u = s_v$, we wish to turn $u$ and $v$ into twins.
    To accomplish this, we will describe three sets of edges $F_1, F_2, F_3 \subseteq E$ to delete. 

    \paragraph{The cell tree $T$.}
    We begin by describing a ``segment tree''-type structure which we will use to guide our definitions of our deleted edges $F_1, F_2,$ and $F_3$.
    Suppose without loss of generality that $n$ is a power of two.
    This can be achieved by adding dummy vertices with no edges, which do not affect the rank-width or number of edges of the resulting graph.
    
    Let $T$ be the complete rooted binary tree with $2 \e^{-1} n$ leaves. 
    We then associate the number $i$ with the $i$th leaf of $T$ in a fixed depth-first-search order, so that there is a bijection between $\{0, 1, \ldots 2 \e^{-1} n - 1\}$ and the leaves of $T$.
    We then inductively, in a bottom-up fashion, associate each non-leaf vertex of $T$ with the union of its children.
    That is, every level of $T$ partitions $\{0, 1, \ldots, 2 \e^{-1} n - 1\}$ into contiguous subsequences. 
    We will henceforth refer to the vertices of $T$ as \EMPH{cells} and say that a vertex of $T$ associated with a set of size $2^i$ is a $2^i$-cell. 
    Note that each $2^i$-cell $C$ is associated with a set of the form $\{k \cdot 2^i, k \cdot 2^i + 1, \ldots, (k+1) \cdot 2^i - 1\}$ for some $k \in \mathbb N$.
    We will occasionally abuse notation to write $C = [k \cdot 2^i, (k+1) \cdot 2^i)$.

    Let $\theta \in \{0, 1, \ldots, \e^{-1} n -1\}$ be a uniformly sampled integer from $[0, \e^{-1} n)$.
    For a vertex $v \in V$, we say that $v$ is contained in a cell $C = [k \cdot 2^i, (k+1) \cdot 2^i)$, and write $v \in C$, if $k \cdot 2^i \leq \theta + v < (k+1) \cdot 2^i$.
    In the expression $\theta + v$, we are interpreting $v$ as its position in the left-to-right ordering of $V$. 
    If $s_v \neq 0$, we define the \EMPH{home cell} of $v$ as $C_v$, the unique $(\e^{-1} s_v)$-cell containing $v$.
    If $s_v = 0$, we let $C_v$ be the unique $1$-cell containing $v$.
    We will see towards the end of this section that the tree $T$, with some modifications, will later be the basis of our rank-decomposition of $G$.
    In this tree, a vertex's position will be roughly determined by its home cell.

    \paragraph{$F_1$: Restricting forward neighborhoods to home cells.}
    We now define our first set of deleted edges $F_1$. 
    For $v \in V$ and $u \in N^+(v)$, we add the edge $vu$ to $F_1$ if $u \not \in C_v$, the home cell of $v$. 
    That is, set
    \[
        F_1 := \{vu \in E \mid v < u \text{ and } u \not \in C_v\}.
    \]

    \begin{lemma}
        We have that $\mathbb E[|F_1|] \leq 2 \e |E|$.
    \end{lemma}

    \begin{proof}
        Fix any $v \in V$ with $s_v \neq 0$. 
        Let $\theta_{\e^{-1} s_v} \in \{0, 1, \ldots, \e^{-1} s_v - 1\}$ be the residue of $\theta$ modulo $\e^{-1} s_v$.
        Because $\theta$ was selected uniformly at random from $\{0, 1, \ldots, \e^{-1} n - 1\}$, we have that $\theta_{\e^{-1} s_v}$ is uniformly distributed over $\{0, 1, \ldots, \e^{-1} s_v - 1\}$.
        This implies that $v$ occupies a uniformly random ``position'' within its home cell $C_v$.
        Thus, the probability that $F_1$ contains exactly $i$ edges from $v$'s forward neighborhood, for $i \in \{1, 2, \ldots, d^+_v\}$, is precisely $\frac{1}{\e^{-1} s_v}$.
        We can bound the expected number of edges deleted from $v$'s forward neighborhood as
        \[
            \sum_{i=1}^{d_v^+} \frac{i}{\e^{-1} s_v} \leq \frac{{d_v^+}^2}{\e^{-1} s_v} < 2 \e d_v^+. 
        \]
        The lemma follows from summing over all $v \in V$ and the identity $|E| = \sum_{v \in V} d_v^+$.
    \end{proof}

    \paragraph{$F_2$: Normalizing forward neighborhoods within home cells.}
    We now define our second set of deleted edges $F_2$.
    Fix some $v \in V$ with $s_v > \e^{-1}$.
    We aim to coarsify $N^+(v)$ to be a union of $\e s_v$-cells.
    For $u \in N^+(v)$ contained in $\e s_v$-cell $C$, we add the edge $vu$ to $F_2$ if not all of $C$ is contained in $N^+(v)$.
    For $v \in V$ with $s_v \leq \e^{-1}$, we do not add any edge from the forward neighborhood of $v$ to $F_2$, as its forward neighborhood is already a union of $1$-cells.

    \begin{lemma}
        We have that $|F_2| \leq 2 \e |E|$.
    \end{lemma}
    \begin{proof}
        Fix any $v \in V$ with $s_v > \e^{-1}$.
        Consider any $u \in N^+(v)$ with $vu \in F_2$.
        Then $u$ is either contained in the first or last cell of $v$'s forward neighborhood.
        Thus, the number of edges in $F_2$ from $v$'s forward neighborhood is at most $2 \e s_v \leq 2 \e d_v^+$.
        The lemma follows from summing over all $v \in V$ and the identity $|E| = \sum_{v \in V} d_v^+$.
    \end{proof}

    \paragraph{$F_3$: Normalizing scales of backwards neighborhoods.}

    The deletions $F_1$ and $F_2$ both serve to normalize the forward neighborhoods of vertices. 
    We now define $F_3$, which helps us get a handle on backwards neighborhoods.
    For vertices $u, v \in V$ with $u < v$, we delete the edge $uv$ if the scale of $u$ is substantially smaller than that of $v$: $s_u \leq \e s_v$.
    Formally, we define
    \[
        F_3 := \{uv \in E \mid u < v \text{ and } s_u \leq \e s_v\}.
    \]

    \begin{lemma}
        We have that $|F_3| \leq 2 \e |E|$.
    \end{lemma}
    \begin{proof}
        Fix any $v \in V$. We will charge all deleted edges $uv \in E$ with $u < v$ to $d^+_v$. 
        Consider any such $u \in V$.
        Then $v$ must be in $u$'s forward neighborhood, $v \in N^+(u)$, so $v \in [u+1, u + d^+_u]$.
        We have that $d^+_u < 2 s_u \leq 2 \e s_v$. 
        Thus, we must have that $u$ is in the range $[v - 2 \e s_v, v - 1]$. 
        This range has $2 \e s_v \leq 2 \e d^+_v$ vertices contained within it, so there can be at most $2 \e d^+_v$ edges in $F_3$ of this form.
        The lemma follows from summing over all $v \in V$ and the identity $|E| = \sum_{v \in V} d^+_v$.
    \end{proof}

    Letting $F := F_1 \cup F_2 \cup F_3$, we have that $\mathbb E[|F|] \leq 6 \e |E|$. 
    To de-randomize, we can simply select the $\theta \in \{0, 1, \ldots, \e^{-1} n - 1\}$ which minimizes $|F|$.
    Thus, we can assume that $|F| \leq 6 \e |E|$.
    Set $G' := (V, E \setminus F)$. 
    It remains to bound the rank-width of $G'$.

    \subsection{Bounding Rank-Width}

    We will bound the rank-width of $G'$ by constructing a rank-decomposition with width $\mathcal O( \e^{-5} \log \e^{-1})$.

    \paragraph{Rank*-decompositions.} Fully specifying a rank-decomposition for $G'$ is somewhat tedious.
    We instead define an alternate rank-decomposition and show that it can be turned into a typical rank-decomposition in a black box manner.
    Given a sub-cubic tree $T'$ and an assignment of vertices $V$ to nodes of $T'$, we say that $T'$ is a \EMPH{rank*-decomposition} of $G'$ with width $(Q, r)$ if
    \begin{enumerate}
        \item each node of $T'$ has at most $Q$ vertices assigned to it and
        \item the implied cut $S_e \subseteq V$ corresponding to each edge $e \in E(T')$ has rank $\rho(S_e) \leq r$.
    \end{enumerate}

    Here, the implied cut $S_e \subseteq V$ corresponding to $e \in E(T')$ consists of all vertices assigned to nodes on one particular side of the edge $e$.

    \begin{lemma} \label{lem:rankstartorank}
        If a graph $H$ has a rank*-decomposition with width $(Q, r)$, then $H$ has rank-width at most $Q + r$.
    \end{lemma}
    \begin{proof}
        Let $T^*$ be a rank*-decomposition of $H$ with width $(Q, r)$. 
        We will construct a rank-decomposition $T$ of $H$ with width at most $Q + r$.
        To construct $T$, we attach an additional subtree to each node of $T^*$.
        If the node of $T^*$ has $\ell \leq Q$ vertices assigned to it, we attach an arbitrary subtree with exactly $\ell$ leaves, and arbitrarily assign those $\ell$ vertices to the leaves. 
        Note that this tree may have nodes of degree $4$.
        To make $T$ sub-cubic, we simply split any node of degree $4$. 
        Each of those duplicates gets $2$ of $T$'s original edge, and the duplicates themselves are attached by an edge. 
        
        We claim that $T$ has width at most $Q + r$.
        To show this, we will consider three types of edges in $T$.
        First, consider a cut defined by an edge $e \in E(T)$ contained in one of our added subtrees.
        Then one side of this cut has at most $Q$ vertices and so has rank at most $Q$.
        Next, consider a cut defined by an edge $e \in E(T)$ which is also present in $T^*$.
        Then this cut is also implied by $T^*$ and so has rank at most $r$.
        The only cuts remaining are those defined by an edge $e \in E(T)$ which we introduced when splitting nodes of degree $4$.
        These cuts differ from cuts implied by $T^*$ by at most $Q$ vertices. 
        Moving a vertex across a cut increases the rank by at most $1$, so their rank is bounded by $Q + r$.
    \end{proof}

    \paragraph{Removing twins.} As a pre-processing step, as long as there exist \emph{false twins} of $G'$, i.e., distinct $u, v \in V$ such that $N(u) = N(v)$, we will delete one of $u$ or $v$.

    \begin{lemma} \label{lem:killtwins}
        Let the graph $H'$ be obtained from another graph $H$ by deleting one of a pair of false twins $u, v \in V(H)$.
        Then the rank-width of $H$ is at most the rank-width of $H'$.
    \end{lemma}
    \begin{proof}
        Let $r$ be the rank-width of $H'$ and $T'$ be a rank-decomposition of $H'$ with width $r$. 
        Let $u$ be the vertex deleted from $H$ to obtain $H'$.
        We construct a rank-decomposition $T$ for $H$ from $T'$ as follows.
        Let $w$ be the leaf in $T'$ assigned to $v$.
        We add two vertices to $T$, each adjacent only to $w$, assigning one to $u$ and the other to $v$.

        Now we claim that $T$ has width $r$. 
        Consider any cut $S_e \subseteq V(H)$ implied by an edge $e \in E(T)$.
        Suppose $e$ is an edge in $T'$ and let $S_e' \subseteq V(H')$ be the implied cut in $H'$.
        Then, letting $S_e'$ be the side of the cut containing $v$, we have that $S_e = S_e' \cup \{u\}$. 
        Let $A_e$ be the cut matrix for $S_e$ and $A_e'$ be the cut matrix for $S_e'$.
        Then $A_e$ is the same as $A_e'$ but with one additional row which is identical to a row already contained in $A_e'$.
        This cannot increase the rank, so we have that $\operatorname{rank}(A_e) = \operatorname{rank}(A_e') \leq r$.
        The only two remaining cuts to consider are those on the two edges we introduced in $T$.
        Each of these cuts has only one vertex on one side of the cut and so have rank at most $1 \leq r$. 
        Thus, $T$ has width $r$ and so $H$ has rank-width at most $r$.
    \end{proof}

    We proceed by assuming $G'$ has no false twins, as we can add them back in at the end without increasing rank-width due to \Cref{lem:killtwins}.

    \paragraph{Bounding $Q$.}
    We now define a rank*-decomposition for $G'$.
    Recall that our cell tree $T$ is defined as the complete rooted binary tree with $2 \e^{-1} n$ leaves.
    Our rank*-decomposition is just this tree $T$ with each vertex $v \in V$ assigned to its home cell $C_v$. 
    We must first bound $Q$, the number of vertices assigned to each cell.
    For a vertex $v \in V$ with scale $s_v$, define $\ell_v$ as the $\max\{\e^2 s_v, 1\}$-cell of $v$ and $r_v$ as the final $\max\{\e s_v, 1\}$-cell of $v$'s forward neighborhood after deleting $F_1 \cup F_2$ from $G$. 

    \begin{lemma}
        If two vertices $u, v \in V$ share the same home cell $C$ and have $\ell_u = \ell_v$ and $r_u = r_v$, then $u$ and $v$ are false twins in $G'$.
    \end{lemma}
    \begin{proof}
        Without loss of generality, suppose $u < v$.
        Because $u$ and $v$ have the same home cell, we have that $s_u = s_v$.
        Further, we have that $\e^2 s_v > 1$ because they share the same $\e^2 s_v$-cell.
        Because $u$ and $v$ share the same $\e s_v$-cell, the edge $uv$ is deleted by $F_2$, so $u$ and $v$ are not adjacent in $G'$.
        Fix any $w \in V \setminus \{u, v\}$. We will show that either $w$ is adjacent to both of $u$ and $v$, or neither, with cases based on the relative ordering of $w$ to $u$ and $v$.

        In the first case, we have that $u < v < w$. Then $w$ is adjacent to $v$ if and only if $s_v > \e s_w$, and $w$'s $\e s_v$-cell lies strictly after the $\e s_v$-cell of $v$, and before the end of $r_v$.
        These conditions are exactly equivalent for $u$ because $(\ell_u, r_u) = (\ell_v, r_v)$.
        Thus, $w$ is adjacent to $v$ if and only if $w$ is adjacent to $u$.

        In the second case, we have that $u < w < v$. Here, we have that $w$ is in the same $\e s_u$-cell as $u$, so $u$ is not adjacent to $w$ due to $F_2$.
        If $s_w \leq \e s_v$, then $w$ is not adjacent to $v$ due to $F_3$.
        If $s_w > \e s_v$, then $\e s_w > \e^2 s_v$.
        Notably, $w$ and $v$ share the same $\e^2 s_v$ cell, so they share the same $\e s_w$ cell, and thus they are not adjacent due to $F_2$.

        In the third case, we have that $w < u < v$. 
        If $s_w \leq \e s_v$, then $w$ is adjacent to neither $u$ nor $v$ due to $F_3$.
        If $s_w > \e s_v$, then $\e s_w > \e^2 s_v$, so $u$ and $v$ share the same $\e s_w$-cell.
        Since the forward adjacency of $w$ is entirely based on scale and $\e s_w$-cells, and $u$ and $v$ share the same $\e s_w$-cell, we have that $w$ is adjacent to $u$ if and only if $w$ is adjacent to $v$.
    \end{proof}

    \begin{corollary} \label{corr:homecellsmall}
        If $G'$ has no false twins, then there are at most $\e^{-5}$ vertices with $C$ as their home cell for any cell $C$.
    \end{corollary}
    \begin{proof}
        If there are no false twins, then $C$ can only contain one vertex per choice of $(\ell_v, r_v)$.
        There are $\e^{-3}$ choices for $\ell_v$ and at most $\e^{-2}$ choices for $r_v$, yielding the desired bound.
    \end{proof}

    Thus, we get the bound $Q \leq \e^{-5}$ for our rank*-decomposition of $G'$.

    \paragraph{Bounding $r$.}
    We now bound $r$, the rank of any cut implied by $T$.

    \begin{lemma} \label{lem:boundr}
        For any $e \in E(T)$, the cut $S_e \subseteq V$ implied by $e$ has $\rho_{G'}(S_e) \leq \mathcal O(\e^{-5} \log \e^{-1})$.
    \end{lemma}
    \begin{proof}
        Let $C$ be the cell immediately below the edge $e$.
        Then $S_e$ is the set of vertices whose home cell is contained within $C$; we write $V_C = S_e$ to emphasize this.
        We will bound the number of different neighborhoods that $u \in V \setminus V_C$ can have within $S_e$.
        Fix any $u \in V \setminus V_C$ and let $C_u$ denote the home cell of $u$.
        If $C_u \cap C = \emptyset$, then $u$ is not adjacent to any of $S_e$ due to $F_1$.
        Note that the cells form a laminar family, so the only remaining case is $C \subseteq C_u$.

        Let us first consider the case where $|C_u| \geq \e^{-2} |C|$.
        Then we have that $s_u = \e |C_u|  \geq \e^{-1} |C|$. 
        We will show that $u$ is either adjacent to all of $V_C$ or none of $V_C$, based on the position of $u$ compared to $C$.

        In the first case, $u < v$ for all $v \in C$. 
        Recall that the forward neighborhood of $u$ is determined by $\e s_u \geq |C|$-cells. 
        Also, because $s_u \geq |C|$, no edges from $u$ to $V_C$ will be deleted by $F_3$.
        Thus, $u$'s adjacency to $V_C$ is determined entirely based on $|C|$-cells.
        Since all of $V_C$ is in one $|C|$-cell, either $u$ is adjacent to all or none of $V_C$.

        In the second case, $u \in C$.
        For any $v \in V_C$ with $v < u$, we have that $s_v \leq \e |C| \leq \e s_u$, and so $v$ is not adjacent to $u$ due to $F_3$.
        For any $v \in V_C$ with $u < v$, we have that $u$ and $v$ share the same $\e s_u$-cell, so $v$ is not adjacent to $u$ due to $F_2$.
        So in this case, $u$ has no edges to $V_C$.

        In the final case, $u > v$ for all $v \in C$.
        Then, as above, for any $v \in V_C$, we have that $s_v \leq \e |C| \leq \e s_u$, and so $v$ is not adjacent to $u$ due to $F_3$.
        So in this case, $u$ has no edges to $V_C$.

        Now we must go back and consider the case where $|C_u| < \e^{-2} |C|$.
        In this case, there is nothing we can do - it is possible that $u$'s neighborhood with $V_C$ is unstructured.
        However, there are only $\mathcal O(\log \e^{-1})$ such cells $C_u$, each of which contains only $\mathcal O(\e^{-5})$ vertices after removing false twins due to \Cref{corr:homecellsmall}.
        Thus, even if each has an entirely different neighborhood, the total number is still bounded by $\mathcal O(\e^{-5} \log \e^{-1})$.
        The additional one from vertices with $|C_u| \geq \e^{-2} |C|$ is absorbed by the constant factor.
        Thus, the cut matrix of $V_C$ has at most $\mathcal O(\e^{-5} \log \e^{-1})$ distinct columns and so has rank bounded above by $\mathcal O(\e^{-5} \log \e^{-1})$.
    \end{proof}

    \begin{corollary}
        $G'$ has rank-width at most $\mathcal O(\e^{-5} \log \e^{-1})$.
    \end{corollary}

    \begin{proof}
        By \Cref{corr:homecellsmall} and \Cref{lem:boundr}, $G'$ has a rank*-decomposition of width at most $(\e^{-5}, \mathcal O(\e^{-5} \log \e^{-1}))$.
        Applying \Cref{lem:rankstartorank} yields an upper bound of $\mathcal O(\e^{-5} \log \e^{-1})$ on the rank-width of $G'$.
    \end{proof}

    Recall that we deleted at most $6 \e |E|$ edges to obtain $G'$. 
    Linearly re-scaling $\e$ finishes the proof of \Cref{thm:structural}.

    \section{AI Disclosure}
    All proofs in this paper were found through the authors' interactions with generative AI tools.
    In particular, a QPTAS for \csp on bounded rank-width graphs was found through discussions with ChatGPT-5.6 Sol, which was improved to a PTAS by Claude Fable 5. 
    Following this, the authors were able to prove a weak structural statement --- that interval graphs are $\e |E|$ edge deletions from having $\mathcal O(\log n)$ rank-width. 
    This bound was then improved to $\mathcal O(1)$ through discussions with ChatGPT-5.6 Sol.
    
    The paper itself is human-written, with some AI assistance in producing figures, polishing presentation, and proofreading for accuracy and typos. 
    All proofs have been simplified and verified by the authors, who take full responsibility for the contents of the paper.

\bibliography{references}
\bibliographystyle{alpha}

\end{document}